\documentclass[letter]{aa} % for the letters 
\usepackage{graphicx}
\usepackage{txfonts}
\usepackage{natbib}
\bibpunct{(}{)}{;}{a}{}{,} % to follow the A&A style

\begin{document}

	\title{First detection of CO lines in a water fountain star.}

	\author{J.~H. He\inst{1,2}
			\and H. Imai\inst{3}
			\and T.~I. Hasegawa\inst{1}
			\and S.~W. Campbell\inst{1,4}
			\and J. Nakashima\inst{5}}

\institute{Institute of Astronomy and Astrophysics, Academia Sinica, P.O. Box 23-141, Taipei 10617, Taiwan (Contact: J.H. He)\\ 
				\email: jhhe@asiaa.sinica.edu.tw
			\and
				National Astronomical Observatories/Yunnan Observatory, Chinese Academy of Sciences, PO Box 110, Kunming, Yunnan Province 650011, PR China
			\and
				Department of Physics, Faculty of Science, Kagoshima University, 1-21-35 Korimoto, Kagoshima 890-0065, Japan
			\and
		     Centre for Stellar and Planetary Astrophysics, School of Mathematical Sciences, Monash University, Melbourne, Australia 3800
			\and
				Department of Physics, University of Hong Kong, Pokfulam Road, Hong Kong
          }

   \date{Received September 15, 1996; accepted March 16, 1997}

% \abstract{}{}{}{}{} 
% 5 {} token are mandatory
 
\abstract
{Water fountain stars are very young post-AGB stars with high velocity water maser jets. They are 
the best objects to study the onset of bipolar jets from evolved stars due to their young 
dynamical ages. However, none of them have been observed in any thermal lines.}
{Search for CO lines in the water fountain star IRAS\,16342-3814 and investigate 
the properties of its thermal gas.}
{The proximity, peculiar stellar velocity and high Galactic latitude of IRAS\,16342-3814 make a single dish 
observation possible. We use the Arizona Radio Observatory 10m telescope to observe the CO
\,J=2$-$1 line and compare the line parameters with that of masers.}
{We report the detection of $^{12}$CO and $^{13}$CO\,J=2$-$1 lines from IRAS\,16342-3814.
The inferred $^{12}$CO mass loss rate is
an order of magnitude lower than the infrared and OH mass loss rates,
indicating a very cold and thick O-rich circumstellar envelope around the star. 
We also find a $^{12}$CO expansion velocity of 
$V_{\rm exp}=46\pm 1$\,km\,s$^{-1}$ that is too high for an AGB wind 
and confirm the systemic velocity of $44\pm1$ km/s. In addition we measure a very low
$^{12}$CO/$^{13}$CO line ratio of 1.7.}
{The first detection of CO lines has provided a new way to
    investigate the water fountain stars.
Given the high expansion velocity of the CO gas and its relation to maser velocities, 
we infer that 
the CO emission region is co-located with the OH mainline masers in the warm base of
the optical bipolar lobes, while the high velocity OH\,1612\,MHz and H$_2$O masers are
located in the side walls and at the farthest ends of the bipolar lobes, respectively.
Further observations are highly desired to understand the very low $^{12}$CO/$^{13}$CO line ratio.}


   \maketitle
%
%________________________________________________________________

\section{Introduction}

Water fountain stars are a group of rare objects that display a thick circumstellar 
envelope (CSE) that is typical for asymptotic giant branch (AGB) stars and also  
high velocity bipolar jets ($\gtrsim 100$\,km\,s$^{-1}$) detected via their H$_2$O masers. 
They could be stars near the end of AGB or post-AGB evolution stage when
non-spherical circumstellar structures begin to develop inside the relic of the 
spherical AGB circumstellar envelope. Currently only about 10 such objects 
have been discovered through single dish and interferometry observations of their OH, H$_2$O and SiO 
masers~\citep[see the review by][]{Imai07a}. However, none of them have ever been 
detected in a thermal molecular line. In this letter, we report the first 
detection of the $J=2-1$ transition of $^{12}$CO and $^{13}$CO in the water fountain 
star \object{IRAS\,16342-3814}. 

\object{IRAS\,16342-3814} was discovered by \citet{Likk88} to show H$_2$O 
masers with a line-of-sight expansion velocity up to $117.4$\,km\,s$^{-1}$,
much higher than the typical speed of $\sim
15$\,km\,s$^{-1}$ for a spherical AGB wind \citep[see, e.g.,][]{Chen01}. 
They also found a progressive 
increase of expansion velocity in OH masers, from 36\,km\,s$^{-1}$ (1665\,MHz) through 
42\,km\,s$^{-1}$ (1667\,MHz) to 58\,km\,s$^{-1}$ (1612\,MHz). 
The symmetrical H$_2$O maser velocities found by \citet{Likk92} allowed 
them to determine a star velocity of $V_{\rm sys}=43.2\pm 0.9$\,km\,s$^{-1}$, which 
further constrained a higher line-of-sight H$_2$O maser expansion velocity of $130$\,km\,s$^{-1}$.

Optical and infrared imaging to this star has revealed striking
structures in its CSE. 
The Hubble Space Telescope (HST) observations by \citet{Saha99} revealed a bipolar nebula. 
Corkscrew patterns were found in the bipolar lobes in follow-up
infrared $L_{\rm p}$ band imaging 
with the Keck telescope \citep{Saha05}, indicating that the bipolar lobes and 
the H$_2$O masers might be created by precessing jets. Infrared imaging 
in the $3.8\sim 20\mu$ range by \citet{Dijk03} 
showed the variation of the CSE morphologies from bipolar at near infrared to spherical at middle 
infrared, demonstrating that the outer CSE ejected earlier was spherical and the star 
is currently undergoing CSE morphology transformation. \citet{Dijk03} also found from the $2\sim 200\mu$ Infrared 
Space Observatory (ISO) spectra that the star is the only object known to show absorption in both
amorphous and crystalline silicate features. The very red spectral energy distribution
possibly indicates a very cold and thick dusty CSE with an AGB mass loss
rate on the order of $10^{-3}\,M_{\odot}{\rm yr}^{-1}$ \citep{Dijk03}.

Interferometric measurements of masers have supplied kinematic
information. The mapping of OH masers in
\object{IRAS\,16342-3814} by \citet{Saha99} and 
\citet{Zijl01} showed linear increase of line-of-sight 
outflow velocity with projected distance from the 
central star. High-velocity OH maser spots are 
located around the optical bipolar 
lobes while low-velocity maser spots appear near 
a dark waist close to the central star. These facts 
are in accord with a wind interaction model of the bipolar 
structure \citep{Zijl01}. However, the corkscrew pattern 
in the infrared images also suggests that precessing jets may play some role.
Very Long Baseline Array (VLBA) mapping of H$_2$O 
masers by \citet{Morr03} and \citet{Clau04} showed bipolar distribution 
of maser spot clusters with more than $2\arcsec$ separation 
along the optical bipolar lobes and quasi-linear alignment of maser spots that 
delineates shock front of the jets. The interplay 
of the jets with magnetic field is still unclear, 
although extremely high degree of polarisation 
has been found in the OH masers \citep{TeLi88,Szym04}.

The post-AGB nature of \object{IRAS\,16342-3814} 
has been critically investigated by \citet{Saha99}. 
However the status of thermal gas in the CSE of this star 
is still unclear due to the lack of thermal 
molecular line observations. Unlike other known water 
fountain stars, \object{IRAS\,16342-3814} is the easiest object in which 
to search for its CO lines with a single dish telescope, because it 
is relatively near with a distance of $0.7\sim2$\,kpc \citep{Saha99,Zijl01},
it has a high galactic latitude of 
$5.^\circ8$ and a large line-of-sight velocity difference of at least
40\,km\,s$^{-1}$ from that of nearby interstellar molecular clouds. 

\section{Observation and data reduction}

The ALMA band-6 sideband separating receiver on the Arizona Radio 
Observatory 10\,m submillimeter telescope (SMT)
was used for the $^{12}$CO and $^{13}{\rm CO}$\,J=2$-$1 observations towards 
\object{IRAS\,16342-3814} on 2008, May 09. Sky subtraction was made 
in a beam switch mode with 2
arcmin throw at 2.2 Hz in azimuth direction. The receiver temperature was about
$100-120$\,K. Two filter banks (FFBs,
1\,GHz width, 1024 channels) were used for the two polarisations of the $^{12}$CO line 
and two acousto-optical spectrometers (AOS's, 972.8\,MHz width, 2048 channels) 
were used for the two polarisations of the $^{13}$CO line. 
A 90 minute integration under a $\tau_{225}=0.15$ and a $T_{\rm sys} = 445$\,K at a low 
elevation of $19^\circ$ resulted in an RMS noise of about
10\,mK at a resolution of 3\,MHz in each polarisation. Main beam efficiencies of 0.7 and 0.6 were used
to convert the measured $T_{\rm A}^*$ into main beam temperature $T_{\rm
  mb}$ for the USB ($^{12}$CO) and LSB ($^{13}$CO), respectively. A telescope
efficiency of 31 Jy/K was used to derive line peak flux. The beam width is 
about $32\arcsec$ at both lines.

The GILDAS/CLASS package was used to reduce and analyse the data. The two
polarisations were combined to improve the S/N. 
A zero order baseline was removed from
each spectrum. We also experimented with a higher order baseline, 
but no much difference was found at the band center where 
the CO lines appear. The SHELL method in the CLASS package was used to fit
both lines with a truncated parabolic line profile.

\section{Results}

\begin{figure*}[hb]
\centering
\includegraphics[scale=.5,angle=270]{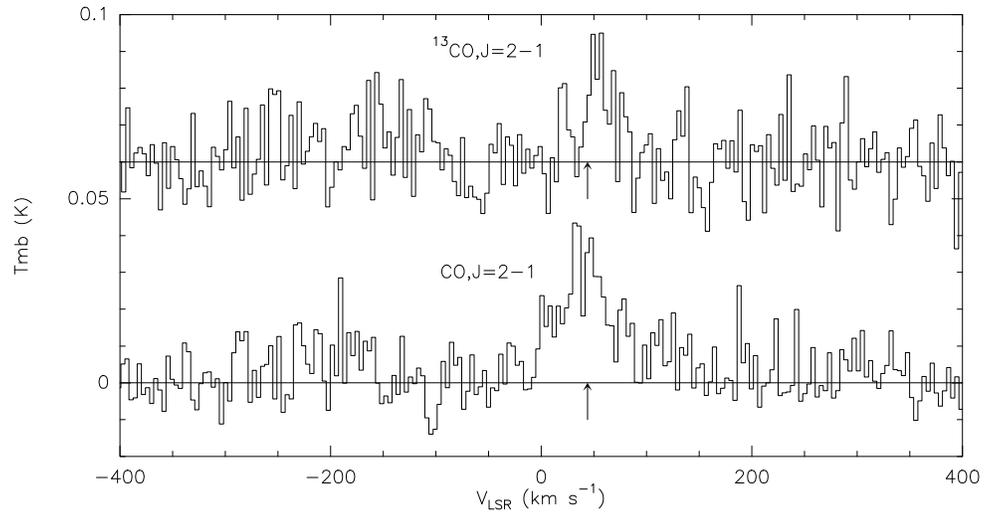}
\caption{The observed $^{12}$CO and $^{13}$CO\,2$-$1 line profiles. The spectra have been 
  box smoothed to a resolution of about 3\,MHz. The
  $^{13}$CO\,2$-$1 line has been shifted upward by 0.06\,K for clarity. The arrows
  mark $V_{\rm sys}=44$\,km\,s$^{-1}$ that is derived from the $^{12}$CO line. 
  \label{f1}}
\end{figure*}

The resulting spectra of $^{12}$CO and $^{13}$CO\,J=2$-$1 lines are shown in Fig.~\ref{f1}, where each
spectrum has been box smoothed to a resolution of $\sim 3$\,MHz. We do not see any
strong interstellar emission or absorption signatures. 
Line profile fitting of $^{12}$CO line gives a peak temperature of $T_{\rm mb}=29$\,mK 
(with ${\rm rms}=7$\,mK), corresponding to a peak flux of 0.90\,Jy (with ${\rm rms}=0.22$\,Jy). The
line area is $2.14\pm 0.14$\,K\,km\,s$^{-1}$,
representing a $15\sigma$ detection of the line. 
A systemic velocity of $V_{\rm sys}(^{12}{\rm CO})=44\pm 1$\,km\,s$^{-1}$ and a high expansion
velocity of $V_{\rm exp}(^{12}{\rm CO})=46\pm 1$\,km\,s$^{-1}$ are measured by line profile
fitting. The $V_{\rm sys}(^{12}{\rm CO})$ agrees well with the H$_2$O maser systemic 
velocity of $43.2\pm 0.9$\,km\,s$^{-1}$ from \citet{Likk92}.
We also notice a pair
of narrow features near the centre of the $^{12}$CO\,2$-$1 line
profile. However, because of the poor S/N, we are not sure if they are real
features. 

The $^{13}$CO\,2$-$1 spectrum is more noisy. We find from line profile fitting a line peak
temperature of $T_{\rm mb}(^{13}{\rm CO})=17$\,mK (with ${\rm rms}=9$\,mK), 
corresponding to a peak flux of 0.53\,Jy  (with ${\rm rms}=0.28$\,Jy). The
line area of $1.02\pm 0.17$\,K\,km\,s$^{-1}$
represents a $6\sigma$ detection of the line. The stellar velocity and
CSE expansion velocity 
found from line profile fitting are $V_{\rm sys}(^{13}{\rm CO})=50\pm 3$\,km\,s$^{-1}$ and
$V_{\rm exp}(^{13}{\rm CO})=34\pm 2$\,km\,s$^{-1}$.
The velocity differences between the $^{12}$CO and $^{13}$CO\,2$-$1 lines
could be merely caused by underestimated noise in the spectra.

\section{Discussion}

\subsection{The low $^{12}$CO mass loss rate}
\label{mlr}

The detection of the $^{12}$CO\,2$-$1 line allows us to directly estimate the mass loss rate 
of \object{IRAS\,16342-3814}. Using the $^{12}$CO\,2$-$1 formula scaled by 
\citet{Hesk90} from a $^{12}$CO\,1-0 formula postulated by \citet{Knap85}, 
we derive $\dot{M}(^{12}{\rm CO}) = 0.7\sim 5.6\times 10^{-6}M_{\odot}{\rm yr}^{-1}$ 
for a source distance of $0.7\sim 2$\,kpc. As discussed in \S~\ref{vel}, the detected CO gas may not 
represent the spherical AGB wind. The true AGB wind CO mass loss rate, 
if also estimated with Heske's formula, should be even 
lower than this value, because the CO gas emission from the slow AGB wind is not definitely detected 
yet, while the mass loss rate is $\propto V_{\rm exp}^2 T_{\rm mb}(CO)$. However, the formula of 
\citet{Baud83} applied to an OH\,1612\,MHz maser peak flux of 9\,Jy \citep{Likk88} 
yields a much higher mass loss rate of 
$\dot{M} {\rm (OH)} = 1.7\sim 4.9\times 10^{-5}M_{\odot}{\rm yr}^{-1}$ 
(with $V_{\rm exp}=46$\,km\,s$^{-1}$ from our $^{12}$CO line). 
\citet{Do07} fitted the Spitzer MIPS $70\mu$ images with a 
two shell model, and obtained
$\dot{M} (70\mu) = 0.4\sim 3\times 10^{-5}M_{\odot}{\rm yr}^{-1}$ for an inner 
smooth CSE and
$\dot{M} ({\rm shell}) = 0.4\sim 3\times 10^{-4}M_{\odot}{\rm yr}^{-1}$ for an outer 
extended dust shell (assuming a gas-to-dust mass ratio of 200 and 
$V_{\rm exp}=15$\,km\,s$^{-1}$. Here we note that our measured $^{12}$CO expansion 
velocity of $46$\,km\,s$^{-1}$ may not be that of the dusty CSE, 
see the discussion in \S~\ref{vel}). The inner shell mass loss rate agrees with the 
OH mass loss rate. Due to its huge radius of $1.4\sim4$\,pc, the outer shell could be explained 
by sweep-up of interstellar material, as suggested 
by the recent hydrodynamic simulation of \citet{Ware07}.
To summarise, we suggest that \object{IRAS\,16342-3814}, while located in a very
extended and cold detached outer dust shell, 
 has an inner CSE whose mass loss rate 
(estimated from IR and OH emission) is roughly an order of magnitude higher 
than its $^{12}$CO mass loss rate. The weakness of $^{12}$CO\,2$-$1 line could be interpreted by 
the coldness of the CO gas, as suggested for some
other very red OH/IR stars by \citet{Hesk90} .

\subsection{The high $^{12}$CO expansion velocity}
\label{vel}

The $^{12}$CO expansion velocity $V_{\rm exp}(^{12}{\rm CO})=46$\,km\,s$^{-1}$ is 
too high for a
typical OH/IR star \citep[typically $\sim 15\pm 5$\,km\,s$^{-1}$ for long
period OH/IR stars, see e.g.,][]{Chen01}. Another possible source of the broad
$^{12}$CO emission is the swept-up bipolar bubble walls. According to 
a momentum driven wind-interaction model 
suggested by \citet{Zijl01}, the radial motion velocity of different parts of the bubble wall will be 
proportional to distance to the central star. This has been confirmed by the linear 
OH maser velocity offset-radius (V-r) relationship found by \citet{Saha99} and \citet{Zijl01}.
Assuming that $V_{\rm exp}(^{12}{\rm CO})$ follows the same V-r relationship in Fig.~3 of \citet{Saha99},
we find a rough angular radius of $0.5\arcsec$ for the CO emission region, similar to that of 
OH mainline masers. 
Further assuming a Gaussian source shape, a brightness temperature of the $^{12}$CO\,2$-$1 line 
can be estimated to be $\sim 30$\,K. Therefore, the $^{12}$CO\,2$-$1 line could be optically thin
if the gas temperature is much higher than this.

Further insight can be gained by comparing the velocities of 
$^{12}$CO\,2$-$1 line, OH and H$_2$O masers. We find several interesting facts 
by comparing the CO velocities with the V-r relationship of OH masers in the Fig.~3 of \citet{Saha99}:
1) The 1612\,MHz maser velocity offsets are all 
$\gtrsim V_{\rm exp}(^{12}{\rm CO})$, while most of the OH\,1665/1667\,MHz (mainline) 
maser velocity offsets $\lesssim V_{\rm exp}(^{12}{\rm CO})$;
2) The OH\,1612\,MHz masers show two components -- a pair of bipolar high 
velocity clumps that
tightly follow the linear V-r relation and a peculiar group of points with 
$V_{\rm LSR}\sim 0$\,km\,s$^{-1}$ but scattered in a
large radius range $(-1.1\arcsec\sim 0\arcsec)$;
3) The position of the bipolar 1612\,MHz maser component is not symmetrical with respect 
to the star position, perhaps due to 3D projection effect.
The peculiar group of 1612\,MHz masers
are not only scattered in
radius, but also scattered in position angle with respect to the central star 
in the VLA maps \citep[see Fig.~1b and Fig.~9 of][respectively]{Saha99,Zijl01}. 
This is difficult to explain with a
spherical expanding shell model in which masers should show an elliptical
relation between line-of-sight velocity and projected radial
distance. Taking away this peculiar 1612\,MHz maser component, one can see 
a better V-r relationship among the rest 1612, 1665 and 1667\,MHz masers.
If we further take a velocity span of $\sim 248$\,km\,s$^{-1}$ and 
a spatial separation of $\sim 3\arcsec$ of H$_2$O masers from \citet{Morr03}
and overlap their symmetry center with that of OH masers, 
we can see that the  H$_2$O masers 
follow the same linear V-r relationship quite well.
Upon these arguments, we
can sketch a picture for the CSE as follows: 

Both the OH 
mainline masers and the CO
emission arise from the same region near the base of the bipolar
outflow, the two clumps of high velocity OH 1612 maser spots are
located on the farther and faster part of the bipolar lobe side walls, the 
H$_2$O masers are stimulated by the shock fronts at the two farthest and fastest 
ends of the bipolar lobes, while the
peculiar group of spatially scattered 1612\,MHz maser spots are located
in an independent structure that is blue-shifted by
$\sim -46$\,km\,s$^{-1}$ with respect to the star but not necessarily
associated with the bipolar outflow. The dark waist and the slow AGB wind might be 
simply too cold to be detected in our CO\,2$-$1 line spectra. Actually, the pair of 
faint spikes on the top of the broad $^{12}$CO\,2$-$1 line profile
near the line centre in Fig.~\ref{f1} could be the emission
from the slow AGB wind. If this is true, 
we may estimate a crude AGB wind speed of $\sim 13$\,km\,s$^{-1}$, typical for an AGB star.

\subsection{On the origin of the extremely low $^{12}$CO/$^{13}$CO ratio}

The detection of the $^{13}$CO~J=2$-$1 line yields an extremely small
$^{12}$CO/$^{13}$CO peak flux ratio of 1.7.
A range of $^{12}$CO/$^{13}$CO ratios have been found in PPNe and PNe by other
researchers. \citet{Hase05} concluded from a survey of published $^{13}$CO
observations of PNe \citep{Pall00,Bals02,Bach97,Joss03} that the
$^{12}$CO/$^{13}$CO ratios are about $10\sim 30$ in most PNe, but are $2\sim
3$ in a few and $>60$ in a few others. Some examples of low ratios are:
$^{12}{\rm CO}/^{13}{\rm CO}\sim 4.6$ in \object{AFGL\,618}, $\sim 2.2$ in
\object{M\,1-16} \citep{Bals02}, $\sim 3$ in the PN \object{NGC\,6302}
\citep{Hase03}, and $\sim 3.2$ in \object{HD\,17982} \citep{Joss01}.
\citet{Bach97} pointed out that the $^{12}$CO/$^{13}$CO ratios in PPNe and
PNe are affected by two competing effects: selective photodissociation of
$^{13}$CO and isotopic fractionation of $^{13}$C in CO. Another issue is the opacity 
in the $^{12}$CO lines, e.g., the very low ratio in \object{NGC\,6302} was 
later found by interferometry work of \citet{Pere07} to be affected by opacity 
effects in a massive torus and the true $^{12}$C/$^{13}$C should be $>15$. For an
embedded dusty post-AGB star like \object{IRAS\,16342-3814}, we do not need
to worry about the selective photodissociation in the central region. If
the detected CO emission does come from the accelerated bipolar bubble
walls (see the discussion in \S~\ref{vel}), neither isotopic fractionation
nor $^{12}$CO line opacity should be an issue, since the dynamical age of
the bipolar jets is short ($\sim 100$\,yr) and the high expansion velocity 
and high radial velocity gradient (according to wind interaction model) 
might render the 
$^{12}$CO emission not too optically thick. A rough but conservative calculation
of the optical depth under an LVG approximation gives a value smaller than unity. 
However, better CO line data is required to confirm this.
We also find that our $^{12}$CO/$^{13}$CO ratio is about 1.1 and 3.2 on the red and blue 
line wings ($46>|V_{\rm exp}|>15$\,km\,s$^{-1}$) respectively, 
but the uncertainties are larger there. One more issue 
is that the $^{12}$CO and $^{13}$CO lines might come from different 
clumps of gas. Again, better CO line profiles and, better still, CO mapping data 
are needed to clarify this point and thus to constrain a reliable $^{12}$C/$^{13}$C ratio. 
Therefore, later in this section, we will discuss different opacity cases separately, 
using only the assumption that
the $^{12}$CO and $^{13}$CO lines are co-located in the CSE.

A precise measurement of the $^{12}$C/$^{13}$C ratio has the potential to cast light on
the evolutionary status of the star. Stellar models of low
mass ($\textrm{M}\lesssim 2.0 \textrm{M}_\odot$) and intermediate mass
($2.0\textrm{M}_\odot\lesssim\textrm{M}\lesssim 8.0 \textrm{M}_\odot$) stars
\citep[e.g.,][]{Elei94,Stra97,Kara07} show that there are quite a few
phases of evolution where this ratio is altered from the initial value
(assumed to be solar at 89). The first occurs during the first dredge up
(FDU) at the beginning of the Red Giant Branch (RGB).  Here the ratio
declines from the initial value to $\sim 20$ as the convective envelope
moves in and mixes up partially burnt material 
(we note that the second dredge-up
has only a minor effect on the ratio). 
The second evolutionary phase that alters the $^{12}$C/$^{13}$C ratio is not
predicted by stellar models. Observations have shown that the isotopic ratio
reduces along the RGB in low-mass stars ($\textrm{M} \lesssim 2.0
\textrm{M}_\odot$), via a process known as Deep Mixing \citep[also known as
$\delta_\mu$ mixing - see the recent theoretical models by][]{Eggl08}. This
phenomenon reduces the ratio down to a value of $\sim 15$ by the end of the RGB in
Population I stars.
The third phase that significantly alters the
ratio actually increases it -- the Third Dredge-Up (TDU). Here $^{12}$C is
periodically mixed into the envelope during the thermally-pulsing AGB stage
(TPAGB), raising the
$^{12}$C/$^{13}$C ratio from the FDU value to values of $\sim60$ or more
\citep[see e.g.,][for recent observations of this effect]{Lebz08}.
The final phase to alter the ratio is Hot Bottom Burning (HBB). However this
only occurs in stars more massive than $\sim 4$ or $5 M_\odot$ (at solar
metallicity). HBB is active during the later phases of the TPAGB and acts
to burn the carbon and nitrogen in the convective envelope via the CNO
cycles. This results in the $^{12}$C/$^{13}$C ratio approaching the equilibrium
value of $\sim 3$ \citep[see e.g.,][for some observational evidence of
this phenomenon]{Plez93,Mcsa07}. The very low $^{12}$C/$^{13}$C ratio in some Galactic PNe, PPNs
and extremely young post-AGB stars like \object{IRAS\,16342-3814} could be also
indicative of the HBB process.

If \object{IRAS\,16342-3814} has a high $^{12}$C/$^{13}$C ratio close to the solar value of 89
(i.e., $\sim50$ times larger than 
the $^{12}$CO/$^{13}$CO line ratio, e.g., as a 3\,M$_\odot$ star experiencing TDU), 
the CO line optical depths can 
be derived from the observed $^{12}$CO/$^{13}$CO ratio (=1.7) to be 
$\tau(^{12}{\rm CO})\sim82$ and $\tau(^{13}{\rm CO})\sim0.93$ (according to $T\propto(1-e^{-\tau})$ 
under the assumption of co-location of $^{12}$CO and $^{13}$CO gas and the same excitation temperature). 
However, the $^{12}$CO mass loss rate 
should also be raised by a factor of $\tau/(1-e^{-\tau})=82$ to $5.7\sim45.9\times10^{-5}{\rm M}_\odot$/yr, 
which is significantly higher than the OH and IR mass loss rates.

If the true $^{12}$C/$^{13}$C ratio is an intermediate value of 20, i.e., 
a factor of $\sim10$ higher than the $^{12}$CO/$^{13}$CO line ratio, the similarly 
determined opacities are $\tau(^{12}{\rm CO})\sim19$ and $\tau(^{13}{\rm CO})\sim0.93$. 
In this case, the $^{12}$CO mass loss rate is comparable to the OH and
IR mass loss rates. 
Therefore, a low mass AGB star ($\lesssim1.5$\,M$_\odot$) that didn't
experience TDU or HBB and thus retains its 
FDU plus Deep Mixing surface abundances would be a possible progenitor of \object{IRAS\,16342-3814}.

If the $^{12}$CO\,2-1 line is only slightly optically thick so that 
the $^{12}$C/$^{13}$C ratio is $\sim3$ (the CNO equilibrium value), the CO line opacities are 
$\tau(^{12}{\rm CO})\sim2.3$ and $\tau(^{13}{\rm CO})\sim0.77$. In this case, we can see that the only process 
(to the best of our knowledge) that could have
lowered the $^{12}$C/$^{13}$C ratio to such a value is HBB. Thus the inferred mass of the star that produced the
material that we observe today would have been $> 4$ or 5 M$_\odot$. 
Interestingly the low
$^{12}$CO/$^{13}$CO ratio, bipolar outflow, and the properties of the OH
masers of \object{IRAS\,16342-3814} are reminiscent of the massive PPN
\object{IRAS\,22036+5306} investigated by \citet{Saha03,Saha06}, although
\object{IRAS\,16342-3814} is probably in an earlier phase of evolution due
to the presence of the young `water fountain'.
Alternatively, a binary system in which a massive AGB star had transfered 
low $^{12}$C/$^{13}$C matter to the current AGB star is also possible.
We note that a low isotopic ratio is supported by our rough estimation
of the small $^{12}$CO\,2-1 opacity at the beginning of this section.

However, both the measured CO and H$_2$O maser systemic velocities 
\citep[$\sim$44\,km\,\,s$^{-1}$,][]{Likk92} confirm that 
the spatial motion of \object{IRAS\,16342-3814} is in the reverse direction of the Galactic
disk rotation \citep[interstellar CO cloud velocity near the same line of sight is 
$\sim -17\,{\rm km\,s}^{-1}$, ][]{Dame01}. The peculiar 
stellar motion lends weight to \object{IRAS\,16342-3814} being a low mass stars, although in rare cases kick-out of stars from 
the crowded galactic plane can give rise to peculiar velocities.

Given the preceeding discussions it is obvious that further observations are highly desirable to 
investigate the different opacity possibilities. Molecular line interferometry and maser proper
motion measurements \citep[such as the work of][]{Imai07b} are also needed to check the 
interstellar contamination, confirm the co-location of the $^{12}$CO and $^{13}$CO gas,
and to locate the birthplace of the possibly escaped AGB progenitor on the Galactic plane.

\begin{acknowledgements}

We thank ARO telescope operators for the assistance in remote
observations. J.H. thanks the projects No. 10433030 and 10503011 of
the National Natural Science Foundation of China. T.I. acknowledges
the support from NSC grant NSC 96-2112-M-001-018-MY3. The SMT is
operated by the Arizona Radio Observatory (ARO), Steward Observatory,
University of Arizona.

\end{acknowledgements}

%{\it Facilities:} \facility{HHT ()}.

\bibliographystyle{aa}  % style aa.bst
\bibliography{ms}   % my bib file

%\clearpage

\end{document}